\documentstyle[12pt,aps,prd,preprint,tighten]{revtex}
\def\Tr{\mathop{\rm Tr}\nolimits}
\def\M{{\cal{M}}}
\def\y{$\times$ }
\def\p{ $+$}
\def\m{$-$ }
\def\Re{\mathop{\rm Re}\nolimits}
\def\Im{\mathop{\rm Im}\nolimits}
\begin{document}
\title{\Huge Constraints on spin observables \\ in 
 $\bar{ p}{ p}  \rightarrow \overline{\Lambda}{\Lambda}$}
\preprint{\normalsize\parbox{4cm}{ISN 99-35\\ hep-ph/9905400\\}}
\author{Mokhtar Elchikh} 
\address{Universit\'e des Sciences et de la Technologie d'Oran,\\
BP 1505 El M'naouer, Oran, Algeria}
\author{Jean-Marc   Richard}
\address{Institut des Sciences Nucl\'eaires,
 Universit\'e Joseph Fourier--CNRS-IN2P3,\\
 53, avenue des Martyrs, 38026 Grenoble Cedex, France}
\date{\today}
\maketitle
\begin{abstract}     
    It is recalled that spin-observables in the strangeness-exchange 
    reaction $\bar{ p}{ p} \rightarrow \overline{\Lambda}{\Lambda}$ 
    are not independent but are related to each other by simple 
    algebraic relations.  This provides constraints on the existing 
    data on polarization and spin-correlation coefficients, and also 
    on the forthcoming data obtained using a polarized proton target.
\end{abstract}


Antihyperon--hyperon production $( \overline{ Y }{ Y})$ in 
antiproton--proton $(\bar{ p}{ p})$ collisions has been studied by the 
PS185 collaboration~\cite{Barnes91,Barnes96} at CERN. Experimental 
data on the integrated cross section, differential distribution $I_0$, 
polarization $P$ and spin-correlation coefficients $C_{ij}$ at various 
energies have already been published.  The experiment has been resumed 
using a polarized proton target and the results of the analysis are 
expected to be published soon~\cite{Johansson99}.

The algebra of observables involved in the scattering of two spin 1/2 
particles is rather straightforward and has been extensively studied 
in dedicated articles~\cite{Bystricky78,LaFrance80,Delaney75}.  This 
knowledge has, however, been somewhat lost, and it seems desirable to 
adapt the general formalism to the special case of the set of 
observables available for $\bar{ p}{ p} \rightarrow 
\overline{\Lambda}{\Lambda}$.

Furthermore, there is considerable interest in the study of this 
reaction.  Models based on ${ K}$, ${ K}^{*}$ 
exchange~\cite{Haidenbauer92a}, quark-pair creation or polarized 
strange sea-quarks in the nucleon~\cite{Alberg95} give different 
predictions on the transfer of spin polarization.  This was the main 
motivation for extending the study of strangeness production.  The 
problems encountered there are intimately related to those of deep 
inelastic scattering or OZI-violation in $\bar{ p}{ p}$ 
annihilation~\cite{Alberg95}.

In Ref.~\cite{JMR96}, it was recalled that the existing data on correlation
coefficients give constraints on the transfer of polarization
 from ${ p}$ to ${\Lambda}$  or from 
 ${ p}$ to $ \overline{\Lambda}$. 
In the present note, we wish to provide further inequalities, which hopefully
 will be  useful for analyzing the data.

We start from the decomposition of the transition matrix $\M$
into 6 (complex) amplitudes $a,b,c,d,e$ and $g$. 
In the c.m.s, $\M$ can be written as~\cite{Bystricky78,LaFrance80}:
\begin{eqnarray} 
\M=&&(a+b)I +(a-b) {\vec \sigma}_1.{\rm\hat{ n}} \, {\vec \sigma}_2 .{\rm\hat{n}}  
 +(c+d) {\vec \sigma}_1. {\rm\hat{k}} \, {\vec \sigma}_2 . {\rm\hat{k}}  \nonumber\\ 
 &&{}+(c-d) {\vec \sigma}_1 .{\rm\hat{p}} \, {\vec \sigma}_2. {\rm\hat{p}}  
+e( {\vec\sigma}_1+{\vec \sigma}_2 ). {\rm\hat{n}} 
+g( {\vec \sigma}_1 .{\rm\hat{k}} \, {\vec \sigma}_2 .{\rm\hat{p}}+  
{\vec \sigma}_1 .{\rm\hat{p}} \, {\vec \sigma}_2. {\rm\hat{k}}),
\end{eqnarray}
where the kinematical unit-vectors are defined from the momentum ${\rm 
\vec{p}_{i}}$ of $\bar{\rm p}$ and ${\rm \vec{p}_{f}}$ of 
$\overline{\Lambda}$:
\begin{equation}
\rm\hat{ p}={   {\rm\vec{p}_{f}}  \over |{\rm\vec{p}_{f}}|  }, \qquad     
\rm\hat{ n}={  {\rm\vec{p}_{i}} \times  {\rm\vec{p}_{f}}
            \over |{\rm\vec{p}_{i}}\times {\rm\vec{p}_{f}}|    },  
  \qquad \rm\hat{k}=\rm\hat{n} \times \rm\hat{p}. 
\end{equation}
Neglecting an overall flux and phase-space factor,
 the differential cross-section $ I_{0}$ and 
the spin-observables are given by:
\begin {eqnarray} 
  I_{0}&&=\Tr[\M \M^{\dag}],  \nonumber \\
 P_{n}I_{0}&&=\Tr[\vec{\sigma}_1.\rm\hat{n} \M \M^{\dag}],  \nonumber \\
A_{n}I_{0}&&=
         \Tr[\M {\vec{\sigma}}_{2}.\rm\hat{n} \M^{\dag}], \nonumber \\
C_{ij}I_{0}&&=
     \Tr[\vec{\sigma}_1. \hat{\imath}\,\vec{\sigma}_2 .
           \hat{\jmath} \M \M^{\dag}],\\
D_{ij}I_{0}&&=
       \Tr[\vec{\sigma}_2. \hat{\imath} \M  {\vec{\sigma}}_2 .
           \hat{\jmath} \M^{\dag}],\nonumber \\    
K_{ij}I_{0}&&=
  \Tr[\vec{\sigma}_1. \hat{\imath} \M {\vec{\sigma}}_2 .
         \hat{\jmath} \M^{\dag}].\nonumber  
\end{eqnarray}
More explicitly (again, up to an overall factor):
\begin{eqnarray} 
\label{Observables} 
 I_0&&=|a|^2+|b|^2+|c|^2+|d|^2+|e|^2+|g|^2,\nonumber \\
 P_{n}I_0&&=2\Re (ae^*)+2\Im (dg^*),\nonumber \\
 A_{n}I_0&&=2\Re (ae^*)-2\Im (dg^*),\nonumber \\
C_{nn}I_0&&=|a|^2-|b|^2-|c|^2+|d|^2+|e|^2+|g|^2,\nonumber \\
C_{xx}I_0&&=-2\Re (ad^*+bc^*)-2\Im (ge^*),\nonumber \\
C_{zz}I_0&&=2\Re (ad^*-bc^*)+2\Im (ge^*),\nonumber \\
C_{xz}I_0&&=-2\Re (ag^*)-2\Im (ed^*),\nonumber \\
D_{nn}I_0&&=|a|^2+|b|^2-|c|^2-|d|^2+|e|^2-|g|^2,\\
D_{xx}I_0&&=2\Re (ab^*+cd^*), \nonumber \\
D_{zz}I_0&&=2\Re (ab^*-cd^*),\nonumber \\
D_{xz}I_0&&=2\Re (cg^*)+2\Im (be^*),\nonumber \\
K_{nn}I_0&&=|a|^2-|b|^2+|c|^2-|d|^2+|e|^2-|g|^2, \nonumber \\
K_{xx}I_0&&=-2\Re (ac^*+bd^*),\nonumber \\
K_{zz}I_0&&=-2\Re (ac^*-bd^*),\nonumber \\
K_{xz}I_0&&=-2\Re (bg^*)+2\Im (ec^*). \nonumber
 \end{eqnarray}
To project out the spins of the particles,
we follow here the usual convention that for $\overline{\Lambda}$, 
$\{\rm\hat{x},\rm\hat{n},\rm\hat{z}\}$ coincides with 
$\{\rm\hat{k},\rm\hat{n},\rm\hat{p}\}$, while for p or $\Lambda$,
the axes $\{\rm\hat{x},\rm\hat{n},\rm\hat{z}\}$ coincides with 
$\{-\rm\hat{k},\rm\hat{n},-\rm\hat{p}\}$. 

 In principle, a polarized target gives access to rank-3 observables 
 of the type:
  \begin{equation}\label{rank-3}
 C_{0\alpha ij}I_{0}=\Tr[\vec{\sigma}_1.\hat{\imath} \, 
 \vec{\sigma}_2.\hat{\jmath} \M \vec{\sigma}_{2} .\hat{\alpha} 
                    \M^{\dag}].
   \end{equation}
For instance, 
 \begin{eqnarray}\label{explicit-rank-3}
 C_{0nzz}I_0&&=2\Re (de^*)-2\Im (ag^*),\nonumber \\
 C_{0nxx}I_0&&=-2\Re (de^*)+2\Im (ag^*),\\
 C_{0nzx}I_0&&=-2\Re (ge^*)-2\Im (ac^*+bd^*),\nonumber 
 \end{eqnarray}
 $ C_{0nnn}$ being equal to $A_{n}$.  It is yet sure, however, whether 
 a statistically significant measurement of these rank-3 observables 
 will be possible from the data accumulated during the last run of the 
 CERN experiment (PS 185/3) \cite{Johansson99}.

There are linear relations among observables.  For instance, it is 
shown in Appendix that
\begin{eqnarray}\label{ineq-lin-1}
2 |A_n|-C_{nn}&&\le 1,\nonumber \\
2 |P_n|-C_{nn}&&\le 1,\\ 
2 |C_{xz}|-C_{nn}&&\le 1.\nonumber 
\end{eqnarray}
From the proof, it becomes clear that similar inequalities exists 
among combinations of observables, one example being
\begin{equation}\label{ineq-lin-2}
    2\vert A_{n}+P_{n}\vert -(K_{nn}+D_{nn})\le 2.
 \end{equation}
If one now concentrates on the set $\{I_0,C_{nn},D_{nn},K_{nn}\}$ 
\begin{eqnarray}
\label{R1}
 I_0&&=|a|^2+|b|^2+|c|^2+|d|^2+|e|^2+|g|^2,\nonumber \\
 C_{nn}I_0&&=|a|^2-|b|^2-|c|^2+|d|^2+|e|^2+|g|^2,\nonumber \\
 D_{nn}I_0&&=|a|^2+|b|^2-|c|^2-|d|^2+|e|^2-|g|^2,\\ 
 K_{nn}I_0&&=|a|^2-|b|^2+|c|^2-|d|^2+|e|^2-|g|^2, \nonumber 
\end{eqnarray}
 one deduces
\begin{eqnarray}
\label{R2}
4(|a|^2+|e|^2)&&=I_0(1+C_{nn}+D_{nn}+K_{nn})\ge 0,\nonumber \\
 4 (|d|^2+|g|^2)&&=I_0(1+C_{nn}-D_{nn}-K_{nn})\ge 0,\nonumber \\ 
 4|c|^2&&=I_0(1-C_{nn}-D_{nn}+K_{nn})\ge 0,\\
 4|b|^2&&=I_0(1-C_{nn}+D_{nn}-K_{nn})\ge 0. \nonumber 
\end{eqnarray} 

With the help of $A_{n}$ and $P_{n}$, the positivity of $|a|^2+|e|^2$ is 
refined into the separate positivity of $ |a \mp e|^2$, and similarly for
$|d|^2+|g|^2$ into $|d \mp i g|^2$. One easily derives
 \begin{eqnarray}
\label{obsPA}
 1+C_{nn}+D_{nn}+K_{nn}+2P_{n}+2A_{n}&&\ge 0, \nonumber \\
 1+C_{nn}+D_{nn}+K_{nn}-2P_{n}-2A_{n}&&\ge 0, \nonumber \\
 1+C_{nn}-D_{nn}-K_{nn}+2P_{n}-2A_{n}&&\ge 0, \\
 1+C_{nn}-D_{nn}-K_{nn}-2P_{n}+2A_{n}&&\ge 0. \nonumber  
 \end{eqnarray}
The variables
\begin{eqnarray}\label{subs1}
&&a'=(a+d)/\sqrt 2,\qquad b'=(b+c)/\sqrt 2,\qquad  e'=(e+ig)/\sqrt 2,\nonumber \\
&&d'=(a-d)/\sqrt 2,\qquad c'=(b-c)/\sqrt 2,\qquad  g'=(e-ig)/\sqrt 2,
\end{eqnarray}
allow one to rewrite the set  $\{I_0,C_{nn},C_{xx},C_{zz}\}$ as 
\begin{eqnarray}
\label{subs2}  
 I_0&&=|a'|^2+|b'|^2+|c'|^2+|d'|^2+|e'|^2+|g'|^2,\nonumber \\  
-C_{xx}I_0&&=|a'|^2+|b'|^2-|c'|^2-|d'|^2-|e'|^2+|g'|^2,\nonumber \\
 C_{nn}I_0&&=|a'|^2-|b'|^2-|c'|^2+|d'|^2+|e'|^2+|g'|^2,\\
 C_{zz}I_0&&=|a'|^2-|b'|^2+|c'|^2-|d'|^2-|e'|^2+|g'|^2,\nonumber  
 \end{eqnarray}
 and so deduce
\begin{eqnarray}
\label{subs3}
4(|a'|^2+|e'|^2)&&=I_0(1+C_{nn}-C_{xx}+C_{zz})\ge 0, \nonumber \\
4(|d'|^2+|g'|^2)&&=I_0(1+C_{nn}+C_{xx}-C_{zz}) \ge 0,\nonumber \\
4|c'|^2&&=I_0(1-C_{nn}+C_{xx}+C_{zz})\ge 0, \\
4|b'|^2&&=I_0(1-C_{nn}-C_{xx}-C_{zz}) \ge 0. \nonumber    
\end{eqnarray} 
Note that the third of the above relations is nothing but the  spin-singlet
fraction 
\begin{equation}
F_0={1\over4}(1+C_{xx}-C_{yy}+C_{zz})={1\over2 I_0}|b-c|^2,
\end{equation}
being positive.  The normalization is such that $F_0=1/4$ in the 
absence of a spin-dependent interaction.

Let us now provide some examples of quadratic inequalities.  In 
Ref.~\cite{JMR96}, it was recalled that Eqs.~(\ref{Observables}) imply
\begin{equation}\label{disk}
  C_{zz}^2+D_{nn}^2\le 1,
\end{equation}
and a number of the similar inequalities.  The proof is given in 
Appendix.  Table~\ref{Binary} summarizes the pairs of rank-1 or rank-2 
observables which satisfy a quadratic relation similar 
to~(\ref{disk}).

Other relations can be written down, involving combinations of more 
than 2 or 3 observables.  For instance, it will be shown below that 
$D_{nn}=K_{nn}$ when $F_0=0$.  This suggests that $D_{nn}$ cannot 
differ too much from $K_{nn}$ when $F_0$ is small.  It can be shown 
that
 \begin{equation}\label{DKF_0}
  \left({D_{nn}-K_{nn}\over 2}\right)^2+(2F_0-1)^2\le 1,
 \end{equation} 
which relates $D_{nn},K_{nn},C_{nn},C_{xx}$ and $C_{zz}$.
As a consequence, $D_{nn}=K_{nn}$ also in the (unphysical) limit
of a pure spin-singlet reaction.
 
The most general method for writing a number of quadratic equalities 
has been given in Refs.~\cite{LaFrance80,Delaney75}.  One can solve 
Eqs.~(\ref{Observables}) and similar for higher-rank observables to 
extract $aa^{*}$, $ab^{*}$, ...  in terms of experimental quantities.  
Then any identity of the type
\begin{equation}
\label{ab}
(ab^{*})\, (cd^{*})=(ad^{*})\, (cb^{*})
\end{equation}
translates into a relation between observables.  This usually involves 
quantities such as $D_{0\alpha \beta \gamma}$ which are hardly 
measurable.  We refer to Ref.~\cite{Delaney75} for more details in the 
case of ${ p}{ p}$ elastic scattering.

{\bf Special cases}.  There are great simplifications in situations 
where one or more amplitudes (or combinations) vanish, in particular:

{\sl 1)} {\underline {Pure spin-triplet}}: $b=c$, and as a consequence
 \begin{equation}\label{pure-t1}
 D_{xx}=-K_{xx},\qquad D_{zz}=-K_{zz},\qquad  D_{xz}=-K_{xz},\qquad  D_{nn}=K_{nn}.
\end{equation}
This is not a surprise.  If the final state contains only components 
which are symmetric under exchange of ${\Lambda}$ and $ 
\overline{\Lambda}$ spins, then the same correlation is expected 
between ${ p}$ and ${\Lambda}$ spins as between ${ p}$ and $ 
\overline{\Lambda}$ spins.  In the case of pure spin-triplet, we also 
have
\begin{equation}
C_{nn}-C_{xx}\ge 0,\quad C_{nn}-C_{zz}\ge 0,\quad C_{xx}+C_{zz}\ge 0,
\end{equation}
as well as several inequalities similar to those of Eq.~(\ref{ineq-lin-1}),
\begin{equation}
2|D_{zz}|+C_{zz}\le 1, \qquad 2|K_{xx}| +C_{xx}\le 1,
\end{equation}
etc., which are listed in Table~\ref{Tab2}.

 {\sl 2)} {\underline {Forward production}}: This case was discussed 
 recently in Ref.~\cite{Pak99}.  In the forward limit, 
 $\theta_{cm}=0$, the transition matrix $\M$ becomes invariant under 
 any rotation around the beam-axis.  In our notation, this means 
 $e=g=0$ and $a-b=c+d$.  As a consequence, the spin parameters are 
 related.  In particular:
 \begin{equation} \label{R3}
 C_{xx}=-C_{nn}, \qquad D_{xx}=D_{nn}, \qquad K_{xx}=-K_{nn},
 \end{equation}
and
 \begin{equation} 
C_{zz}=1-2|b+c|^2/I_0, \quad D_{zz}=1-2|a-b|^2/I_0,\qquad K_{zz}=-1+2|a-c|^2/I_0,     
\end{equation}
implying 
\begin{equation}
K_{zz}-D_{zz}\le 0,\qquad C_{zz}+D_{zz}\ge 0, \qquad {\rm etc.}
\end{equation}
The relation between $C_{xx}$ and $C_{nn}$ was already 
noticed~\cite{Haidenbauer92a}.

{\sl 3)} Furthermore, in the combined case of pure spin-triplet and 
forward production, the expression for the longitudinal 
spin-observables simplifies to
 \begin{equation}
 {1-C_{zz}\over 2}=D_{zz}=-K_{zz}=4|b|^2/ I_{0} \ge 0.
\end{equation}

Within the errors limits, the linear constraints~(\ref{subs3}) and the 
quadratic inequalities of type~(\ref{disk}), in particular those 
between $P_n$ and $C_{ij}$, seem to be satisfied by the 
data~\cite{Barnes96} except at a few points.  Also $C_{xx}\simeq 
-C_{nn}$ is observed in the forward and backward regions.  So, from this 
point of view, the most recent data~\cite{Barnes96} are more 
consistent than the former ones~\cite{Barnes91}.

Let us summarize.  Several spin-observables can be measured for the 
$\bar{ p}{ p} \rightarrow \overline{\Lambda}{\Lambda}$ reaction for 
the weak decay of ${\Lambda}$ (or $\overline{\Lambda}$) gives an 
indication on its spin.  This offers the possibility to test in great 
detail the mechanisms by which strangeness is created.  The experiment 
is however delicate, and its analysis might use the consistency checks 
provided by the linear or quadratic inequalities listed in this note.  
It is hoped that reliable spin observables will help probing the 
mechanisms proposed for this strangeness-exchange reaction.  In 
particular, the hypothesis of a pure spin-triplet 
$\overline{\Lambda}{\Lambda}$ production, suggested by early LEAR data 
on unpolarized targets, can be tested accurately.

\acknowledgments
We thank A.M. Green for comments on the manuscript. 

\appendix
\section* {}

The quadratic inequality (\ref{disk}) can be derived using the vectors
\begin{equation}
\vec{V}_1=(|a|^2-|d|^2,2|ad|) ,\qquad \vec{V}_2=(|b|^2-|c|^2,2|bc|), 
\qquad  \vec{V}_3=(|e|^2-|g|^2,2|eg|)
\end{equation}
with normalization $|\vec{V}_1|=|a|^2+|d|^2$, 
$|\vec{V}_2|=|b|^2+|c|^2$ and $|\vec{V}_3|=|e|^2+|g|^2$.  Given that 
$C_{zz}I_0 \le 2(|ad|+|bc|+|eg|)$, one can deduce
\begin{equation}
    I_0^2(C_{zz}^2 +D_{nn}^2) \le (\vec{V}_1+\vec{V}_2+\vec{V}_3)^2 \le
  (|\vec{V}_1|+|\vec{V}_2|+|\vec{V}_3|)^2
\end{equation}
and thus the desired $C_{zz}^2 +D_{nn}^2 \le 1$. 
  
Similarly, one can introduce  the vectors
\begin{equation}
 \vec{V}_1=(|a|,|d|), \qquad \vec{V}_2=(|b|,|c|), \qquad \vec{V}_3=(|e|,|g|),
\end{equation}
with normalization $|\vec{V}_1|^2=|a|^2+|d|^2$, 
$|\vec{V}_2|^2=|b|^2+|c|^2$ and $|\vec{V}_3|^2=|e|^2+|g|^2$.  Given 
that $ I_0 A_n \le 2 \vec{V}_1 .  \vec{V}_3$ and $ I_0 D_{xx} \le 2 
\vec{V}_1 .  \vec{V}_2$, one can deduce
\begin{equation}
I_0^2(A_n^2+D_{xx}^2)\le 4|\vec{V}_1|^2(|\vec{V}_2|^2+|\vec{V}_3|^2)\le
 4 I_0^2 (|\vec{V}_1|^2/I_0)[1-(|\vec{V}_1|^2/I_0)] \le I_0^2,
\end{equation}
and thus $A_n^2 +D_{xx}^2 \le 1$. 

To prove the inequality between $C_{zz}$ and $D_{zz}$ in the case of 
pure spin-triplet, one can start from a simplified problem where the 
amplitudes are real and
\begin{equation}
I_0=a^2+2b^2+d^2,\quad C_{zz}I_0=2(ad-b^2),\quad D_{zz}I_0=b(a-d).
\end{equation}
Then
\begin{eqnarray}
I_0(C_{zz}+2D_{zz})=&&a^2+2b^2+d^2-(a-d)^2-4b^2+4b(a-d)\nonumber \\
=&&I_0-(a-d-2b)^2\le I_0.
\end{eqnarray}
It is easily shown that restoring the complex character of the 
amplitudes and possible non-vanishing of $e$ and $g$ cannot do 
anything but strengthen the inequality.  A similar reasoning holds for 
$C_{zz}-2D_{zz}$.  The proof is analogous for the other inequalities 
in Table~\ref{Tab2} and Eq.~(\ref{ineq-lin-1}).


%
\newpage
\begin{table}[hbt]
    \renewcommand{\arraystretch}{1.3} 
    \caption{\label{Binary} Pairs of observables fulfilling an inequality 
    such as $C_{zz}^{2}+D_{nn}^{2}\le 1$}
\begin{center}
\begin{tabular}{ c c c c c c c c c c c c c|c}
 $A_{n}$ & $C_{nn}$ & $C_{xx}$ & $C_{zz}$ & $C_{xz}$ &
            $D_{nn}$ & $D_{xx}$ & $D_{zz}$ & $D_{xz}$ &
	    $K_{nn}$ & $K_{xx}$ & $K_{zz}$ & $K_{xz}$&    \\
\hline
    &   &\y   &\y   &\y   &   & \y    &\y &\y &  & \y  &\y   &\y   & $P_{n}$  \\
    &   &   &   &  &         & \y    &\y &\y &  &\y   &\y   & \y  &$A_{n}$ \\
    &   &   &   &   &   &\y   &\y   &\y &   &\y  &\y   & \y  &$C_{nn}$ \\
    &   &   &   &  \y &\y   &    &\y  &   &\y &    &\y   &     &$C_{xx}$ \\
    &   &   &   &\y   &    \y &\y   &  &\y &\y &\y  &     & \y    &$C_{zz}$ \\
    &   &   &   &      &\y   & \y  & \y  &\y       &\y &\y  &\y   & \y  &$C_{xz}$ \\
    &   &   &   &   &      &   &   &       &  &\y   &\y    & \y  &$D_{nn}$ \\
    &   &   &   &   &   &      &   &\y       &\y &    &\y    & \y  &$D_{xx}$ \\
    &   &   &   &   &   &   &      &\y       &\y &\y  &     & \y  &$D_{zz}$ \\
    &   &   &   &   &   &   &   &          &\y &\y  &\y   &     &$D_{xz}$ \\
    &   &   &   &   &   &   &   &   &      &        &     &     &$K_{nn}$ \\
    &   &   &   &   &   &   &   &   &   &           &   &\y   &$K_{xx}$ \\
    &   &   &   &   &   &   &   &   &   &   &            &\y   &$K_{zz}$ \\
\end{tabular}
 \end{center}
\end{table}
\begin{table}[hbt]
\caption{ \label{Tab2} Pairs of observables fulfilling an inequality 
$2|\alpha|\pm\beta\le 1$.  The relations involving 
$(\alpha,\beta)=(P_n,C_{nn})$, $(A_n,C_{nn})$ and $(C_{xz},C_{nn})$ 
are general, as per Eq.~(\protect\ref{ineq-lin-1}), the others are 
specific of a pure spin-triplet reaction.  }
\renewcommand{\arraystretch}{1.3}
\begin{tabular}{c|*{14}{p{25pt}}}
&\multicolumn{14}{c}{
\raise -4pt\hbox{$\overbrace{\kern 400pt}^{\displaystyle \alpha}$}}\\    
             $\beta$& $P_{n\phantom{n}}$&$A_{n\phantom{n}}$ & $C_{nn}$ & $C_{xx}$ 
& $C_{zz}$ &$C_{xz}$ & $D_{nn}$ & $D_{xx}$ & $D_{zz}$ 
& $D_{xz}$ &$K_{nn}$ & $K_{xx}$ & $K_{zz}$ & $K_{xz}$\\
\hline
$C_{nn}$ &\m  &\m  &  &  &  &\m  &\m  &  &  &  &\m  &  &  & \\
 $C_{xx}$ &  &  &  &  &  &  &  &\p  &  &\p  &  &\p  &  &\p \\
 $C_{zz}$ &  &  &  &  &  &  &  &  &\p  &  &  &  &\p  & \\
 $C_{xz}$ &  &  &  &  &  &  &  &  &  &  &  &  &  & \\
\end{tabular}
\end{table}

\begin{thebibliography}{10}
%
\bibitem{Barnes91}
P.D. Barnes {\sl et al.}, Nucl. Phys. A {\bf 526} (1991) 575.
%
\bibitem{Barnes96}
P.D. Barnes {\sl et al.}, Phys. Rev. C {\bf 54} (1996) 2831.
%
\bibitem{Johansson99}
We thank T. Johansson and K. Paschke for useful information about the recent 
developments of the PS185 experiment.

\bibitem{Bystricky78}
J. Bystricky, F. Lehar and P. Winternitz, J. de Physique
  (Paris) {\bf 39} (1978) 1.
%
\bibitem{LaFrance80}
P. La France and P. Winternitz, J. de Physique (Paris) {\bf 41} (1980) 1391.
%
\bibitem{Delaney75}
R.M. Delaney and J.L. Gammel, Phys. Rev. D {\bf 12} (1975) 1978;
C.~Bourrely and J.~Soffer, {\sl ibid.} 2932.
%
 \bibitem{Haidenbauer92a}
 J. Haidenbauer, K. Holinde, V. Mull and J. Speth, Phys. Lett. B {\bf
   291} (1992) 223; Phys. Rev. C {\bf 46} (1992) 2158.
 %
\bibitem{Alberg95}
 M.A. Alberg, J. Ellis and D. Kharzeev, Phys. Lett. B {\bf 356} (1995) 113.
%
\bibitem{JMR96}
J.-M. Richard, Phys. Lett. B {\bf 369} (1996) 358. The statement $A_n=P_n$ in this paper is erroneous.
%
\bibitem{Pak99}
N.K. Pak and M.P. Rekalo, Phys. Lett. B {\bf 450} (1999) 433.
\end{thebibliography}
\end{document}